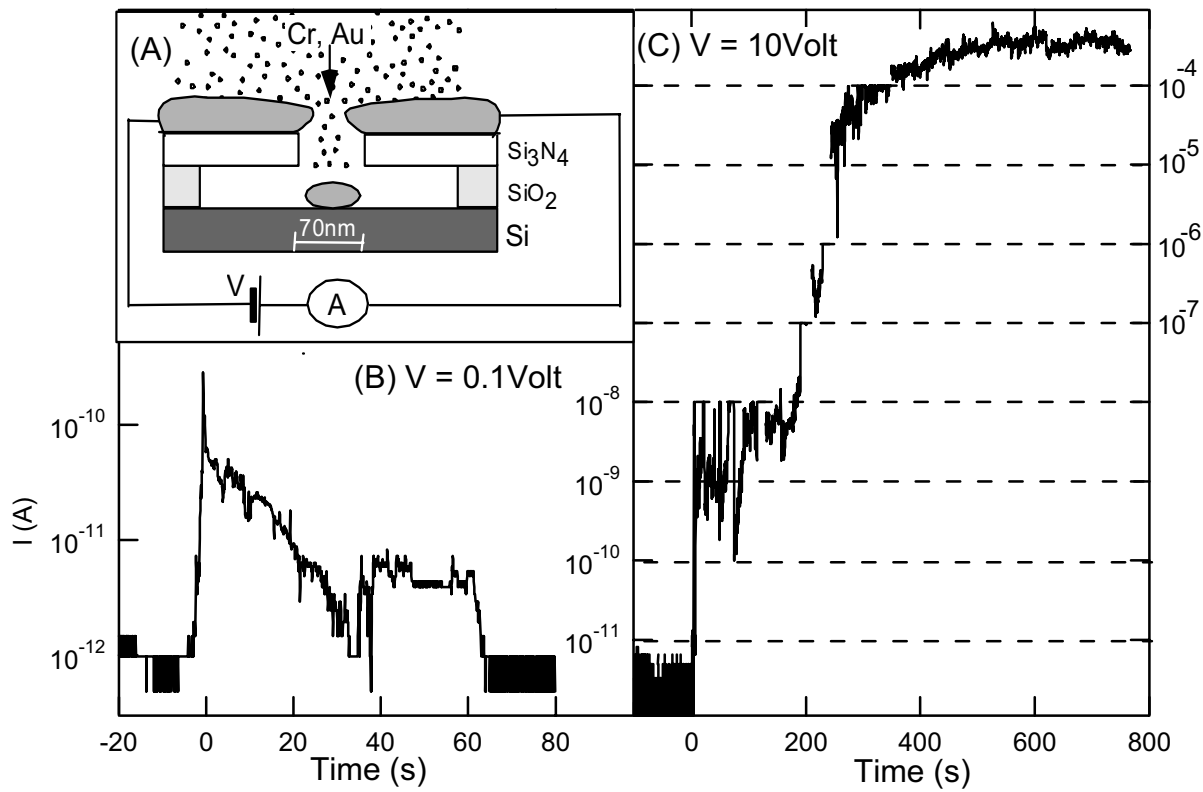

# Electrostatically Assembled Metallic Point Contacts

A. Korotkov, M. Bowman, H. J. McGuinness, and D. Davidovic

School of Physics, Georgia Institute of Technology, Atlanta GA 30332-0430.

We describe a method for creating atomic-scale electric contacts. A metal source is deposited on two insulating substrates separated by a 70 nm gap. Electric conductance across the gap is monitored, while protrusions from the bulk extend into the gap and reach corresponding protrusions on the other side of the gap. When the voltage across the gap is large, a pair of corresponding protrusions assembles into a point contact, via equilibrium between electrostatic attraction and a maximum sustained current. The current voltage characteristics of the point contact display crossover from weak link into tunneling junction when its conductance is near $2e^2/h$, indicating that the point contact is atomic-scale.

Electron transport measurements of single atoms and molecules are fundamentally challenging because they require connection of a molecule to two large-scale conducting electrodes. Early measurements of atoms and molecules were obtained through the use of the scanning tunneling microscope (STM) [1-7]. An STM permits measurement and manipulation of single atoms and molecules on a conducting surface [1, 8-11]. An alternative technique of measuring electron transport in atoms and molecules is through fabrication of sub-nanometer scale electric contacts [12-21]. At small length scale, the sample fabrication is facilitated by self-assembly.

We describe a reproducible fabrication method in which an applied electric field is used to create atomic-scale contacts. The schematic of the experimental setup is shown in Fig. 1(A). First, we create a trench between two insulating silicon-nitride ($Si_3N_4$) substrates, following the recipe in ref. [22]. Fig. 1(A) displays a cross-section of the trench with a 50 nm layer of $Si_3N_4$ on top of a 200 nm layer of dry silicon dioxide ($SiO_2$). Through the $Si_3N_4$ layer, the width of the trench is approximately 70 nm, with a 120 nm wide undercut in the $SiO_2$ layer. Using standard metal vapor deposition techniques with a base pressure in the low $10^{-7}$ Torr range, we deposit a thin metal film, at a rate of 3 Å/s, over an exposed 0.1 mm length of the trench. As we deposit metal atoms over the trench, the undercut serves to prevent connection between the two films. Figure 1(A) also displays that the conductance across the trench is monitored during metal film deposition, and our expectation that the metal film sticks out over the trench.

Before proceeding with fabrication details, we show the main discovery; the applied voltage plays a critical role in the electric contact formation. In initial measurements of current (I) versus time (t), a thick gold film was deposited over the trench, and evaporation was stopped as soon as the slightest current was detected. The resulting I(t) dependence is shown in Fig. 1. If the applied voltage was 0.1 V, the electric current decreased after the initial contact, suggesting that the two metal films that stick out of the trench retracted from each other (Fig. 1(B)). Conversely, if the applied voltage was 10 V, the current increased with time (Fig. 1(C)), suggesting that the two metal films approached each other from electrostatic attraction. The behavior in which the current increases with time will be referred to as electrostatically enhanced growth. The applied voltage is used to opposite effect than in electro-migration [18], where a large applied voltage is used to break a connection between metal films.

We now describe fabrication process in detail. First, we deposit a Cr film with a 45 mV voltage applied across the gap. The voltage is increased to 10V when the thickness of Cr film reaches 20 nm. We stop depositing when the slightest electric current is detected. At this stage, the Cr film thickness varies among samples. Each sample can have a Cr film that varies in thickness from 50 to 70 nm. Samples with a Cr film from 50 to 70 nm have a tunneling resistance in vacuum in significant excess of 1GΩ. In Cr samples we do not observe electrostatically enhanced growth, because Cr is less mobile than Au. In Cr samples, exposure to oxygen, or oxygen in air, decreases the current to below $10^{-11}$ A, due to the formation of Cr surface oxide.

Some of the available Cr samples have been examined by a scanning electron microscope (SEM). In a typical sample, there are typically three to five spots randomly placed along the edge of the trench where, upon close examination with the SEM, protrusions from bulk Cr extend into the trench and reach toward corresponding protrusions on the other side of the trench. One such spot is shown in Fig. 2(A). These spots will be referred to as junctions, and the protrusions will be referred to as electrodes. If we assume that the probability of finding protrusions of different length is distributed by a Gaussian along the 0.1 mm long trench, then it follows that the probability that more than one junction contributes to the current with the same order of magnitude is less than 0.01 (because of the exponential dependence of tunneling conductance on electrode spacing). The Cr junctions serve as templates for subsequent gold evaporation.

Next, an Au film of uniform thickness is deposited onto the Cr template, while the vacuum is maintained in the low $10^{-7}$ Torr range. Among samples, the thickness of Au may vary from 12 to 30 nm, while 10 V is maintained across the gap (trench) during film growth in each sample. The current, exhibiting strong fluctuations, increases by several orders of magnitude after gold deposition. The fluctuations of current in time slowly diminish, and eventually become less than ~20% of the average current, suggesting that the electric contact that forms between gold electrodes reaches an equilibrium. Among samples, the time that it takes to equilibrate varies from several minutes to several hours. The final resistance of the junctions at 10 V bias voltage also varies among samples, from 5kΩ to 50MΩ.

The balance between electrostatic attraction and a maximum sustained current can explain the fluctuations and equilibration in current versus time. Specifically, the applied voltage creates a strong attractive force between gold electrodes, causing the electrodes to approach. The approach reduces the tunneling resistance between electrodes, increasing the current until a maximum sustained current is reached. Then the electrodes suddenly retract, causing the current to decrease. Then, this cycle of electrode attraction and retraction repeats, causing strong current fluctuations. As the cycle repeats, the gold electrodes anneal, becoming more stable against electrostatic attraction. As a result, the fluctuations in current gradually diminish, and equilibrium is ultimately established. The origin of maximum sustained current is not yet understood, and we speculate that it is caused by either electro-migration [18] or by trapping of residual water molecules at the base pressure inside the evaporator [23].

Examination of the samples by an SEM shows there are typically three to five junctions placed randomly along the trench. One such junction is shown in Fig. 2(B). The gold film along the edge has an irregular shape over a distance of approximately 500 nm along the trench. The irregular shape occurs due to structural reconfigurations in response to the applied voltage. The electric contact is at the center of this region. Figure 2(C) shows the junction at higher magnification, indicating that the diameter of the tunneling junctions is less than a few nanometers. Exposure to air at low bias voltage does not change the junction resistance, since gold surfaces do not oxidize. We estimate only one junction contributes significantly to the conductance of a sample, which is corroborated by the observation that the resistance varies strongly among different samples, as described above.

We have studied more than 60 samples. In general, the observations are reproducible between samples. The current-voltage characteristics (I-V curves) at room temperature and at the base pressure of the evaporator are frequency (f) dependent (a saw-tooth voltage at frequency f is applied). The characteristic frequency is approximately $f_0$~1Hz. When f>>$f_0$, the I-V curve of the junction is smooth and reversible. Fig. 3(A) shows a gray-scale image of the probability of different points in the I-V plane, obtained by measuring the I-V curve 4000 times, with corresponding increase in applied voltage. The image shows that the I-V curve is single valued and nonlinear, distinctive of tunneling junctions.

If $f \ll f_0$, the junction conductance exhibits switching during I-V curve measurements, indicating atomic reconfigurations in gold electrodes. By measuring the I-V curve 4000 times, we obtain the probability distribution of I-V points in Fig. 3(B), with corresponding increase in applied voltage. The I-V curve, averaged among different scans, exhibits hysteresis. The standard deviation of junction conductance has a broad maximum between 3 and 6 V. In this voltage range, the atomic reconfigurations are more probable than elsewhere. The probability distribution is such that a continuum of I-V points is more probable than a discrete distribution of points. The switching events essentially vanish below 2 V; at voltages below 1 V the junction resistance varies by less than 10% over a period of a day, indicating that the drift in spacing between the electrodes is less than 0.004 Å/hour.

The shape of the I-V curve differs between linear and nonlinear, depending on the junction resistance, with a transition occurring at roughly 10 k$\Omega$, as shown by the gray-scale image in Fig. 3(B). The I-V curves of junctions with resistance greater than 10 k$\Omega$ are nonlinear, distinctive of tunnel junctions; the I-V curves of junctions with resistance less than 10 k$\Omega$ are linear, distinctive of metallic weak links. A gap in the current distribution separates the tunneling from the weak link regimes. We conclude that one conducting channel dominates tunneling between the leads, since the transition occurs when the resistance is near that of an ideal gold atomic contact (13 k$\Omega$) [12], indicating that the tunneling junctions are atomic-scale. The transition between tunneling and weak link regimes is analogous to the one in mechanically controlled break junctions [12].

In conclusion, we have illustrated an electrostatically enhanced growth process that creates electric contacts between two gold electrodes. SEM images show that the contacts are less than few nanometers in diameter, and yet, transport measurements suggest that one conducting channel dominates conductance, and that the contacts are atomic-scale. The contacts can sustain up to 10 V and are remarkably stable at low applied voltage. Future research will involve electrostatic trapping of small molecules and tunneling spectroscopy of molecular energy levels at cryogenic temperatures.

This work was performed in part at the Cornell Nanofabrication Facility (a member of the National Nanofabrication Users Network), which is supported by the National Science Foundation under Grant ECS-9731293, Cornell University and industrial affiliates. The authors thank Roberto Panepucci, John Treichler, and Alan Bleier from Cornnell Nanofabrication Facility for technical assistance; Center for Microelectronics Research at Georgia-Tech, for substrate preparation; and the Electron Microscopy Center at Georgia-Tech, for access to SEM. This research is supported by the David and Lucile Packard Foundation Grant 2000-13874 and the NSF Grant DMR-0102960.

Figure 1. (A) Deposition of metal films on two $Si_3N_4$ substrates, separated by a 70 nm gap. Large undercut serves to prevent contact between the films. The conductance between the films is monitored *in situ*. (B) Gold film is grown until a slight current is detected, at time 0, which typically occurs when a film thickness of ~80 nm is reached. The graph shows that at low voltage (0.1 V) the current decreases as a function of time. (C) Current versus time at large applied voltage (10 V). The current slowly increases with time, fluctuates, and eventually stabilizes.

Figure 2. (A) A junction formed by depositing Cr at 10 V-applied voltage. (B) An irregularly shaped Au junction formed by adding 12 nm of Au on top of the Cr junction, at 10 V applied voltage. (C) Image of the junction at 400 000 magnification shows that the contact is of nanometer length scale.

Figure 3. (A) Average I-V curve of a junction obtained by sweeping the applied voltage at a frequency of f = 10 Hz for 4000 consecutive cycles. The image shows an abundance (probability) of different I-V values obtained in 4000 I-V scans. (B) Average I-V curve of a junction obtained by sweeping the applied voltage a frequency of f = 0.1 Hz for 4000 consecutive cycles. The gray scale image shows the abundance (probability) of different I-V values obtained in 4000 I-V scans. The probability distribution depends on the direction of voltage change; the probability distribution in this figure corresponds to increasing voltage. The arrow indicates a linear I-V curve of a junction with resistance of 9.3 k$\Omega$, indicating that the junction is a metallic weak link.